\documentclass[aps,prl,twocolumn,showpacs,superscriptaddress,groupedaddress]{revtex4}  
\usepackage{graphicx}  
\usepackage{dcolumn}   
\usepackage{bm}        
\usepackage{amsthm}
\usepackage{amsmath}
\usepackage{amssymb}
\usepackage{amsfonts}
\usepackage{amscd}
\usepackage{ascmac}
\usepackage{enumerate}
\usepackage{bm}
\usepackage{latexsym}

\newtheorem{thm}{Theorem}

\newtheorem{prp}{Proposition}
\newtheorem{lem}{Lemma}



\def\;{{\hspace{0.3ex};\hspace{0.5ex}}}
\def\,{{\hspace{0,3ex},\hspace{0.5ex}}}
\def\({{\hspace{1.2ex}(}}

\def\R{{\mathbb{R}}}
\def\Z{{\mathbb{Z}}}

\def\C{{\mathbb{C}}}
\def\ph{{\varphi}}

\def\dim{{{\rm dim}}}

\begin{document}


\title{Irreversibility of Entanglement Concentration for Pure State}
\author{Wataru Kumagai$^{1,2}$~~~Masahito  Hayashi$^{2,3}$\\
\textit{${}^1$Graduate School of Information Sciences, Tohoku University, Japan \\${}^2$Graduate School of Mathematics, Nagoya University, Japan,\\${}^3$Centre for Quantum Technologies, National University of Singapore, Singapore}}
\date{\today}

\begin{abstract}
For a pure state $\psi$ on a composite system $\mathcal{H}_A\otimes\mathcal{H}_B$, both the entanglement cost $E_C(\psi)$ and the distillable entanglement $E_D(\psi)$ coincide with the von Neumann entropy $H(\mathrm{Tr}_{B}\psi)$. Therefore, the entanglement concentration from the multiple state $\psi^{\otimes n}$ of a pure state $\psi$ to the multiple state $\Phi^{\otimes L_n}$ of the EPR state $\Phi$ seems to be able to be reversibly performed with an asymptotically infinitesimal error when the rate ${L_n}/{n}$ goes to $H(\mathrm{Tr}_{B}\psi)$. In this paper, we show that it is impossible to reversibly perform the entanglement concentration for a multiple pure state even in asymptotic situation. In addition, in the case when we recover the multiple state $\psi^{\otimes M_n}$ after the concentration for $\psi^{\otimes n}$, we evaluate the asymptotic behavior of the loss number $n-M_n$ of $\psi$. This evaluation is thought to be closely related to the entanglement compression in distant parties.



\end{abstract}

\maketitle



%

The entanglement is an essential resource to apply important quantum processing such as the quantum teleportation and the superdense coding. Then, since those protocols often require a suitable entangled quantum state between distant parties, we need a method transforming a given entangled state to a target entangled state. As typical methods, we focus on the entanglement concentration and dilution in this paper. For a pure state $\psi$ on a composite system $\mathcal{H}_A\otimes\mathcal{H}_B$,
the optimal rate $E_D(\psi)$ of the entanglement concentration is called the distillable entanglement, that is, $E_D(\psi)$ is determined by the supremum of the limit of the ratio $\mathrm{lim}\frac{L_n}{n}$ when the multiple state $\Phi^{\otimes L_n}$ of the EPR state $\Phi$ on $\C^2\otimes\C^2$ can be produced from a multiple pure state $\psi^{\otimes n}$ with an asymptotically infinitesimal error. Similarly, the optimal rate $E_C(\psi)$ of the entanglement dilution is called the entanglement cost, that  is, $E_C(\psi)$ is determined by the infimum of the limit of the ratio $\mathrm{lim}\frac{L_n}{n}$ when a multiple pure state $\psi^{\otimes n}$ can be produced from the multiple EPR state $\Phi^{\otimes L_n}$  with an asymptotically infinitesimal error. Those values coincide with each other and are characterized by the von Neumann entropy $H$ as $E_D(\psi)=E_C(\psi)=H(\mathrm{Tr}_{B}\psi)$\cite{BBPS, HHT}. Especially, it is known that the entanglement concentration and dilution with the optimal rate $\lim {L_n}/{n}=H(\mathrm{Tr}_B \psi)$ are realizable with an asymptotically infinitesimal error, respectively \cite{BBPS}. Thus, by using those entanglement concentration and dilution, it seems that we can perform the entanglement concentration and recover the initial state with an asymptotically infinitesimal error, and thus, the entanglement concentration is reversible in the asymptotic situation. In general, for an arbitrary entangled state $\rho$ which is not necessarily pure on a composite system, the asymptotic reversibility of the entanglement distillation for a multiple state $\rho^{\otimes n}$ is defined by the agreement of the distillable entanglement $E_D(\rho)$ and the entanglement cost $E_C(\rho)$ \cite{VC, YHHS}. 

However, it has not been studied whether both the entanglement concentration and the subsequent recovery operation can be performed with an asymptotically infinitesimal error when the distillable entanglement and the entanglement cost coincide for an entangled state. As is well known, in the non-asymptotic case, the general LOCC transformation has been studied intensively \cite{Nie, JP, LS, Vid}, and as the special case, if a pure state which is different from EPR states can be transformed to EPR states by LOCC without a error, then the initial state can not be exactly recovered by LOCC. Similarly, in the asymptotic case, when the distillable entanglement and the entanglement cost differ for a mixed entangled state $\rho$, it is known that we can not perform both the entanglement distillation and the recovery operation for a multiple state $\rho^{\otimes n}$ with an asymptotically infinitesimal error \cite{VC, YHHS}. On the other hand, in the asymptotic case, it had not been properly discussed until now whether we can perform those operations for a multiple state $\psi^{\otimes n}$ of a pure state $\psi$ with an asymptotically infinitesimal error. To investigate the point, we precisely treat the recovery operation that means the operation to reconstitute the initial state $\psi^{\otimes n}$ (or more generally, a multiple state $\psi^{\otimes M_n}$) with some error after the entanglement concentration, and consider the errors for the entanglement concentration and the recovery operation in this paper. In particular, we focus on the asymptotic behavior of the sum of the errors for a pure entangled state and show the incompatibility between the entanglement concentration and the recovery operation, which implies the irreversibility of the entanglement concentration even in the asymptotic case.

As an application, we evaluate the loss of entanglement for the initial state when the entanglement concentration is used as the entanglement compression. In this setting, the entanglement concentration compresses a multiple entangled state $\psi^{\otimes n}$ into a less dimensional quantum system, and the recovery operation decompresses a multiple state $\psi^{\otimes M_n}$ of $\psi$ with a slight error for large $n$ as is shown in FIG.1. As stated above, the entanglement concentration is irreversible, and hence we can not completely reconstitute the initial state $\psi^{\otimes n}$ after the entanglement concentration with an asymptotically infinitesimal error. Then, we investigate how many copies of $\psi$ vanish at the recovery operation after the concentration when small error $\epsilon$ is permitted. In other words, when we reproduce the multiple state $\psi^{\otimes M_n}$ after the concentration for $\psi^{\otimes n}$, we evaluate the asymptotic behavior of the minimal loss number $n-M_n$ of $\psi$ depending on a permissible error $\epsilon$.

\begin{figure}[t]
 \begin{center}
 \hspace*{2em}\includegraphics[width=70mm, height=85mm,angle=-90]{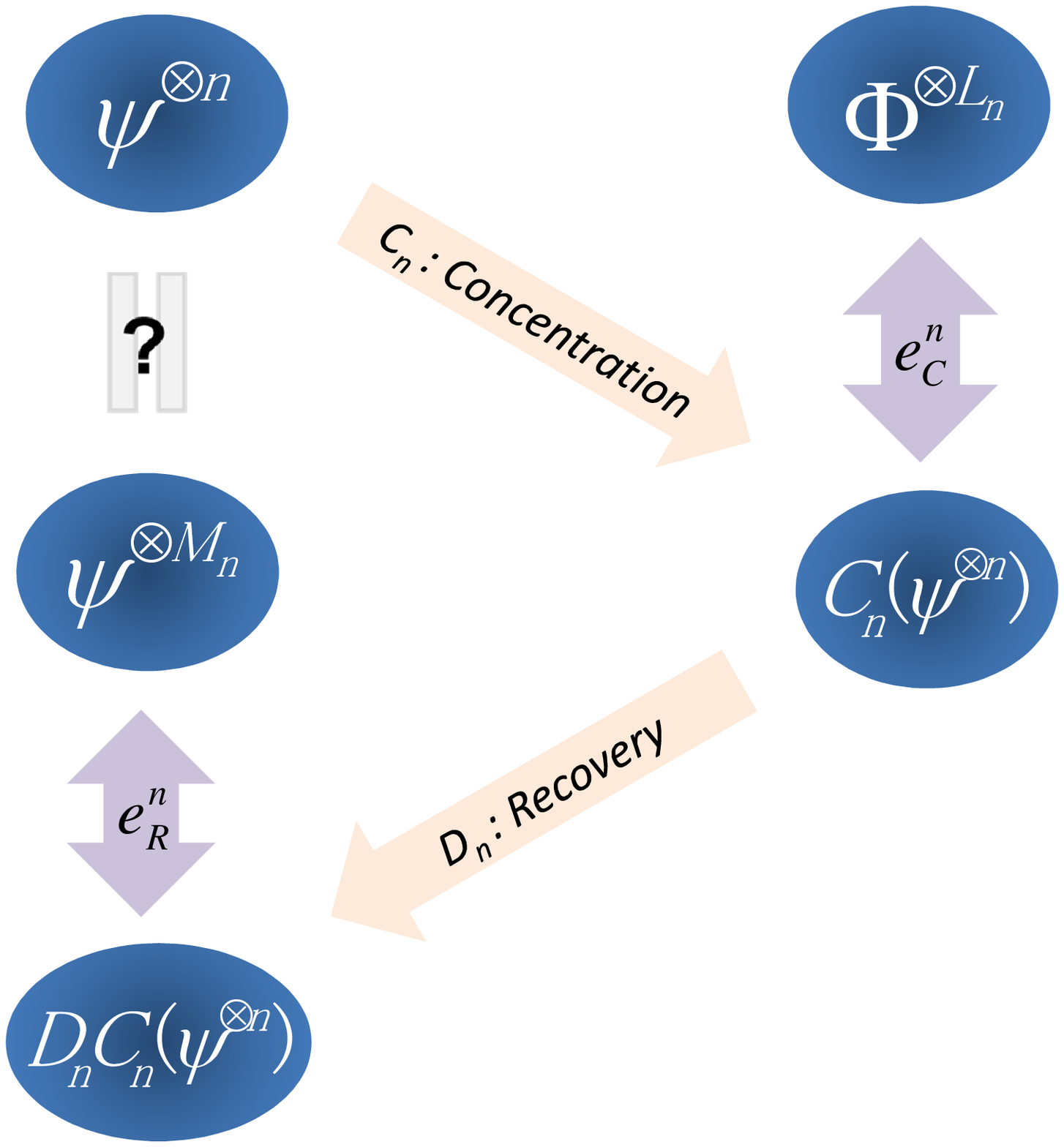}
 \end{center}
 \caption{The entanglement concentration $C_n$ for a multiple pure state $\psi^{\otimes n}$ and the recovery operation $D_n$ of the initial state from the concentrated state $C_n(\psi^{\otimes n})$. $e_C$ represents the concentration error between the target e-bits $\Phi^{\otimes L_n}$ and the concentrated state $C_n(\psi^{\otimes n})$, and $e_R$ does the recovery error between the multiple state $\psi^{\otimes M_n}$ and the restored state $D_n\circ C_n(\psi^{\otimes n})$, respectively.}
 \label{figure2}
\end{figure}


This paper is organized as follows. At first, we introduce an error concerning the entanglement concentration and the recovery operation. Next, we prove the asymptotic incompatibility between the concentration and the recovery, in particular, which implies the irreversibility of the entanglement concentration for an arbitrary pure state $\psi$ except maximally entangled states in the asymptotic situation. Next, we evaluate the asymptotic behavior of the minimum loss number $n-M_n$ of $\psi$ at the recovery. Finally, we summarize our results and give some remarks. For all theorems in this paper, we give the proofs in Supplemental Material.

~


\noindent{\it {Minimum Concentration-Recovery Error:}}~
In the non-asymptotic case, it is known that the entanglement concentration is not reversible \cite{Vid}. On the other hand, in the asymptotic case, it had been thought that the entanglement concentration for a multiple pure state $\psi^{\otimes n}$ is reversible because there exists a pair of the entanglement concentration and dilution protocol with an asymptotically infinitesimal error and with the common optimal rate \cite{BPRST, VC, YHHS,JP}. However, the argument is actually not precise. To clarify the defect of the argument and show the irreversibility of the entanglement concentration in the asymptotic case, we review the operations to implement the entanglement concentration and dilution.

Let $\mathcal{S}(\mathcal{H})$ be the set of all quantum states on a quantum system $\mathcal{H}$. For LOCC transformations $C_n:\mathcal{S}(\mathcal{H}_{A}\otimes\mathcal{H}_{B})^{\otimes n}\to\mathcal{S}(\C^2\otimes\C^2)^{\otimes L_n}$ and $D_n:\mathcal{S}(\C^2\otimes\C^2)^{\otimes L'_n}\to\mathcal{S}(\mathcal{H}_{A}\otimes\mathcal{H}_{B})^{\otimes M_n}$, we call $(L_n,C_n)$ and $(L'_n,M_n,D_n)$ the concentration and dilution map, respectively. In a concentration map, $L_n$ means the number of the e-bits which we generate from the initial state $\psi^{\otimes n}$. Similarly, in a dilution map, $L'_n$ means the number of the e-bits which we want to generate the target state $\psi^{\otimes M_n}$ from. In particular, given a concentration map $(L_n,C_n)$ and a dilution map $(L'_n,M_n,D_n)$, only if $L_n$ equals $L'_n$, we can provide a recovery operation $D_n$ after the entanglement concentration $C_n$. In the situation, we call a quadruplet $(L_n,M_n,C_n,D_n)$ a concentration-recovery map. At the existing discussion about the reversibility of the entanglement concentration \cite{BPRST, VC, YHHS,JP}, only the consistency of the asymptotic rates for the concentration and dilution has been focused and the difference between $L_n$ and $L'_n$ has been missed. In other words, the numbers $L_n$ and $L'_n$ of e-bits have been separately taken under the condition $\lim L_n/n=\lim L'_n/n=H(\mathrm{Tr}_B\psi)$ in the entanglement concentration $C_n$ and dilution $D_n$ although $D_n\circ C_n$ can not be defined. In the following, we treat concentration-recovery maps as pairs of concentration and dilution map with the common number $L_n=L'_n$.

We prepare to introduce an error criteria for a concentration-recovery map. By using the Fidelity $F(\rho,\sigma):=\mathrm{Tr\sqrt{\sqrt{\rho}{\sigma}\sqrt{\rho}}}$, we define the error function $d(\rho,\sigma):={1-F^2(\rho,\sigma)}$. Then we can qualify the entanglement concentration error and the recovery error by
\begin{eqnarray}
&e_C^n(\psi,L_n,C_n):=d(C_n(\psi^{\otimes n}),\Phi^{\otimes L_n}),&\nonumber\\
&e_R^n(\psi,L_n,M_n,C_n,D_n):=d(\psi^{\otimes M_n},D_n\circ C_n(\psi^{\otimes n})),&\nonumber
\end{eqnarray}
respectively. In the following, we focus on the condition 
\begin{eqnarray}\label{error.condition1}
&\lim e_C^n(\psi,L_n,C_n)=0,&
\end{eqnarray}
for a sequence of concentration maps $\{(L_n,C_n)\}_n$. Here, a sequence satisfying $(\ref{error.condition1})$ means the entanglement concentration with an asymptotic infinitesimal error. Similarly, we also focus on the condition
\begin{eqnarray}\label{error.condition2}
&\lim e_R^n(\psi,L_n,n,C_n,D_n)=0&
\end{eqnarray}
for a sequence of concentration-recovery maps $\{(L_n,n,C_n,D_n)\}_n$ when $M_n=n$, and a sequence satisfying $(\ref{error.condition2})$ means the recovery operation with an asymptotic infinitesimal error. Our concern is the compatibility between the entanglement concentration and the recovery operation. In other words, we want to know whether we can carry out both the entanglement concentration and the recovery operation, that is, whether there exists a sequence of concentration-recovery maps satisfying both conditions (\ref{error.condition1}) and (\ref{error.condition2}). Here, let us introduce the significant indicator concerning the compatibility, and call it the minimum concentration-recovery error (MCRE):
\begin{eqnarray}
&&\delta_n(\psi,M_n) \nonumber\\
&&:=\displaystyle\min_{L_n,C_n,D_n} e_C^n(\psi,L_n,C_n)+e_R^n(\psi,L_n,M_n,C_n,D_n)\nonumber
\end{eqnarray} 
where $L_n$ runs over positive integers, and $C_n:\mathcal{S}(\mathcal{H}_A\otimes\mathcal{H}_B)^{\otimes n}\to\mathcal{S}(\C^{2}\otimes\C^{2})^{\otimes L_n}$ and $D_n:\mathcal{S}(\C^{2}\otimes\C^{2})^{\otimes L_n}\to\mathcal{S}(\mathcal{H}_A\otimes\mathcal{H}_B)^{\otimes M_n}$ run over LOCC operations. Then, if there exist the entanglement concentration and the recovery operation simultaneously satisfying $(\ref{error.condition1})$ and $(\ref{error.condition2})$, $\delta_n(\psi,n)$ goes to $0$. Obviously, the converse is correct. Thus, the equation $\lim\delta_n(\psi,n)=0$ corresponds to the compatibility of the entanglement concentration and the recovery with an asymptotically infinitesimal error. In the following, we treat the case of $M_n=n$ and denote $\delta_n(\psi,n)$ simply by $\delta_n(\psi)$.




To evaluate MCRE, we focus on the minimal error concerning the LOCC transformation between $\rho\in\mathcal{S}(\mathcal{H}_A\otimes\mathcal{H}_B)$ and $\sigma\in\mathcal{S}(\mathcal{H}'_{A}\otimes\mathcal{H}'_{B})$:
\begin{eqnarray}
&d(\rho\to\sigma)
:=\displaystyle\min_{E}d(E(\rho),\sigma))
=1-\displaystyle\max_{E}F^2(E(\rho),\sigma),& \nonumber
\end{eqnarray}
where $E:\mathcal{S}(\mathcal{H}_A\otimes\mathcal{H}_B)\to\mathcal{S}(\mathcal{H}'_A\otimes\mathcal{H}'_B)$ runs over LOCC transformation. Then, we get the following equation for $M_n\le n$.
\begin{eqnarray}\label{eq}
&&\delta_n(\psi,M_n)\nonumber\\
&&= \min_{L_n\in\Z_{>0}} d(\psi^{\otimes n}\to\Phi^{\otimes {L_n}}) + d(\Phi^{\otimes {L_n}}\to\psi^{\otimes M_n}).
\end{eqnarray}
The proof is given in Supplemental Material. The equality (\ref{eq}) associates MCRE with the minimal error of the concentration and the dilution, and plays an essential role to show the incompatibility of the entanglement concentration and the recovery operation in the asymptotic situation. To evaluate MCRE, we focus on the right hand side in (\ref{eq}).



~

\noindent{\it {Incompatibility between Entanglement Concentration and Recovery Operation:}}~The entanglement concentration for a multiple pure state $\psi^{\otimes n}$ can be performed with an asymptotically infinitesimal error if the rate $\lim{L_n}/{n}$ of the concentration is less than the distillable entanglement $E_D(\psi)=H(\mathrm{Tr}_B \psi)$ \cite{BBPS}. Moreover, Hayashi et al. \cite{HKMMW} showed the strong converse of the entanglement concentration, that is,  the error $d(\psi^{\otimes n}\to\Phi^{\otimes L_n})$ goes to $1$ if the rate $\lim{L_n}/{n}$ is strictly greater than  $H(\mathrm{Tr}_B \psi)$. Therefore, the asymptotic behavior of the error $d(\psi^{\otimes n}\to\Phi^{\otimes L_n})$ in  (\ref{eq}) is completely analyzed if the rate $\lim{L_n}/{n}$ is not $H(\mathrm{Tr}_B \psi)$. On the other hand, when the rate $\lim{L_n}/{n}$ strictly equals $H(\mathrm{Tr}_B \psi)$, the error $d(\psi^{\otimes n}\to\Phi^{\otimes L_n})$ has not been studied sufficiently. To investigate it, we treat $L_n$ which can be expanded as $L_n=an+b\sqrt{n}+o(\sqrt{n})$, and focus on the coefficients $a$ and $b$, which are called the first and second order rates in information theory, respectively. For $L_n=an+b\sqrt{n}$ and $M_n=n+b'\sqrt{n}$, we get the following theorem.

\begin{thm}\label{2-order}
The equations
\begin{eqnarray}
&&{\lim}d(\psi^{\otimes n} \to \Phi^{\otimes an+b\sqrt{n}})\nonumber\\
&&=\left\{
\begin{array}{ll}
0 & \mathrm{if}~a<H(\mathrm{Tr}_B{\psi}) \\
G\left(\frac{b}{\sqrt{V(\mathrm{Tr}_B{\psi})}}\right) & \mathrm{if}~a=H(\mathrm{Tr}_B{\psi}) \\
1 & \mathrm{if}~a>H(\mathrm{Tr}_B{\psi}),
\end{array}
\right.\\
&&{\lim}d(\Phi^{\otimes an+b\sqrt{n}} \to \psi^{\otimes n+b'\sqrt{n}})\nonumber\\
&&=\left\{
\begin{array}{ll}
1 & \mathrm{if}~a<H(\mathrm{Tr}_B{\psi}) \\
1-G\left(\frac{b-H(\mathrm{Tr}_B\psi)b'}{\sqrt{V(\mathrm{Tr}_B{\psi})}}\right) & \mathrm{if}~a=H(\mathrm{Tr}_B{\psi}) \\
0 & \mathrm{if}~a>H(\mathrm{Tr}_B{\psi}) 
\end{array}
\right.
\end{eqnarray}
hold for any pure state $\psi\in\mathcal{H}_A\otimes\mathcal{H}_B$ except maximally entangled states, where $G(x):=\displaystyle\int_{-\infty}^{x}\frac{1}{\sqrt{2\pi}}\mathrm{e}^{-\frac{x^2}{2}}dx$ and  $V(\rho):=\mathrm{Tr}\rho(-\mathrm{log}\rho-H(\rho))^2$.
\end{thm}
When $\psi$ is a maximally entangled state, $V(\mathrm{Tr}_B{\psi})$ equals $0$ and ${x}/{\sqrt{V(\mathrm{Tr}_B{\psi})}}$ can not be defined in $\R$. However, we can extend Theorem \ref{2-order} for a maximally entangled state by replacing $G(x/\sqrt{V(\mathrm{Tr}_B{\psi})})$ by $0$ or $1$ when $x$ is non-positive or positive, respectively. Theorem \ref{2-order} describes the asymptotic behavior of the errors for the entanglement concentration and dilution. As you can see from the proof, even if $L_n$ has lower order term as $L_n=an+b\sqrt{n}+o(\sqrt{n})$ (e.g. $o(\sqrt{n})=\log n$), the order does not affect the above errors. Hence, when we want to analyze the errors of the entanglement concentration and dilution, we only have to treat the first and second order rate. From Theorem \ref{2-order}, we get the following theorem.





\begin{thm}\label{MCREineq}
 $\displaystyle\underline{\lim}\delta_n(\psi) =1$ holds for any pure state $\psi\in\mathcal{H}_A\otimes\mathcal{H}_B$ except maximally entangled states.
\end{thm}


By Theorem \ref{MCREineq}, far from satisfying $\lim\delta_n(\psi)=0$, MCRE converges to $1$. Therefore, we can not perform both the entanglement concentration and the recovery even if we permit some error $0<\epsilon<1$ for the limit of MCRE as $\overline{\lim}\delta_n(\psi)\le\epsilon$. Here, it turned out that there does not exist a sequence of concentration-recovery maps satisfying both (\ref{error.condition1}) and (\ref{error.condition2}) although there exist a sequence of concentration maps $(L_n,C_n)$ satisfying the condition (\ref{error.condition1}) and a sequence of concentration-recovery maps $(L'_n,n,C'_n,D'_n)$ satisfying the condition (\ref{error.condition2}) with the common first order rates $\lim L_n/n=\lim L'_n/n=H(\mathrm{Tr_B \psi})$. The fact may look strange, but can be comprehended by the argument of the second order rates. That is, those $L_n$ and $L'_n$ actually have different second order rates. The proof at Theorem \ref{MCREineq} is given by using the argument of the second order rate of Theorem \ref{2-order}.

~

\noindent{\it {Loss Evaluation for Recovery Operation:}}~Let us consider the entanglement concentration for the initial state $\psi^{\otimes n}$ and the subsequent recovery operation of the multiple state $\psi^{\otimes M_n}$ satisfying the condition $\overline{\lim}\delta_n(\psi,M_n)\le\epsilon$ for $0<\epsilon<1$. Then, we can not take $M_n$ as $n$ due to Theorem \ref{MCREineq}. If we use the entanglement concentration to compress an entanglement state $\psi^{\otimes n}$ into a less dimensional quantum system, it is significant to know how many copies $n-M_n$ are inevitably lost in the concentration and recovery process. Thus, let us evaluate the rate of loss $n-M_n$ in the asymptotic situation. We focus on the following value for $0<\epsilon<1$ and call it the recovery rate for the entanglement concentration:
\begin{eqnarray}
R(\psi,\epsilon):=\inf_{\{M_n\}}\left\{ \overline{\lim}\frac{n-M_n}{\sqrt{n}}\Big| \overline{\lim}\delta_n(\psi,M_n)\le\epsilon \right\}\nonumber
\end{eqnarray}
The recovery rate means the minimum coefficient of the order of $\sqrt{n}$ of the loss $n-M_n$.
\begin{thm}\label{2nd.rate}
\begin{eqnarray}
R(\psi,\epsilon)=\frac{2\sqrt{V(\mathrm{Tr}_B\psi)}}{H(\mathrm{Tr}_B\psi)}G^{-1}\left(1-\frac{\epsilon}{2}\right)
\end{eqnarray}
\end{thm}
Since $R(\psi,\epsilon)$ is a finite real number for $0<\epsilon<1$, the minimum loss $n-M_n$ of copies can be approximated by $R(\psi,\epsilon)\sqrt{n}$ for large enough $n$ when we perform a suitable entanglement concentration for the initial state $\psi^{\otimes n}$ and recovery operation into the multiple state $\psi^{\otimes M_n}$ with some error $\epsilon$. On the other hand, $R(\psi,\epsilon)$ diverges as $\epsilon$ goes to $0$. Therefore, unlike the case $0<\epsilon<1$, the loss $n-M_n$ increases faster than the order of $\sqrt{n}$ for $\epsilon=0$ in the concentration and recovery process.

~

\noindent{\it {Conclusion:}}~In this paper, we treated the entanglement concentration for a pure state. In existing researches, it has been thought that the initial state can be recovered after the concentration if we perform the concentration with the optimal rate in the asymptotic case. In the argument, the entanglement concentration and dilution have been separately considered although we can not independently perform the concentration and dilution. By simultaneously treating those operations and analyzing the error induced from the concentration and the recovery of the initial state,  it was shown that the sum of the errors is greater than or equal to $1$ as is represented in Theorem \ref{MCREineq}. In particular, when entanglement concentration with an asymptotically infinitesimal error is performed, the recovery error goes to $1$ and it is concluded that the initial state can not be recovered. When we use the entanglement concentration to compress a multiple entangled state $\psi^{\otimes n}$ into a less dimensional quantum system, we derived the asymptotic minimum loss of copies of $\psi$ depending on the permissible error $\epsilon$. As the research relating to the irreversibility of the entanglement concentration, it is conjectured that the LOCC transformation between multiple states of general (pure) states $\psi$ and $\phi$ is irreversible in the asymptotic situation. But it is still an open problem.

We note that the mathematical structures of the entanglement concentration and the recovery operation in quantum information theory are similar to uniform random number generation and source coding in classical information theory. Uniform random number generation treats the way to generate uniform distribution $P^U_{L_n}$ whose support has size $L_n$ by transforming i.i.d. random distribution $P^{\otimes n}$, and source coding does the way to compress data to be able to recover the initial data, respectively. For those problems, it is known that the initial distribution $P^{\otimes n}$ can not be recovered from the transformed distribution after $P^U_{L_n}$ was asymptotically generated by a suitable transformation for $P^{\otimes n}$ when the error is measured by the variational distance \cite{Hay1}. In other words, uniform random number generation is incompatible with source coding. Thus, it can be said that Theorem \ref{MCREineq} corresponds to the incompatibility between uniform random number generation and source coding in classical information theory.




\vspace{0.5em}

\noindent{\it { Acknowledgment:}}~
 WK acknowledges support from Grant-in-Aid for JSPS Fellows No. 233283. MH is partially supported by a MEXT Grant-in-Aid for Scientific Research (A) No. 23246071. 
The Center for Quantum Technologies is funded by the Singapore Ministry of Education and the National Research Foundation as part of the Research Centres of Excellence programme.

\newpage


\begin{center}{\textbf{SUPPLEMENTAL MATERIAL}}
\end{center}

Let us prepare to show the equation (\ref{eq}). For an arbitrary pure state $\psi\in\mathcal{H}_A\otimes\mathcal{H}_B$, we denote the Schmidt coefficients of $\psi$ by $p_{\psi}=(p_{\psi,1},\cdot\cdot\cdot,p_{\psi,M})$. Let $\Phi_{L}=\sum_{i=1}^L \sqrt{1/L}|i\rangle |i\rangle$ be a maximally entangled state with the size $L$ on a quantum system $\mathcal{H}'_A\otimes\mathcal{H}'_B$, which satisfies $L=\min\{\dim\mathcal{H}'_A,\dim\mathcal{H}'_B\}$. When $p^{\downarrow}$ represents the probability distribution which is sorted for the components of a probability distribution $p$ in decreasing order, we define the pure state $\eta_{\psi,L}$ in $\mathcal{H}'_A\otimes\mathcal{H}'_B$ as 
\begin{eqnarray}
\eta_{\psi,L}= 
\displaystyle\sum_{i=1}^{J_{\psi,L}-1}\sqrt{p_{\psi,i}^{\downarrow}}|i\rangle |i\rangle 
+ \sqrt{\frac{\sum_{j=J_{\psi,L}}^M p_{\psi,j}^{\downarrow}}{L+1-J_{\psi,L}}}\displaystyle\sum_{i=J_{\psi,L}}^L|i\rangle |i\rangle\nonumber
\end{eqnarray}
by using
\begin{eqnarray}
J_{\psi,L}:=\max\{1\}\cup\left\{2\le j \le L \Big|\frac{\sum_{i=j}^M p_{\psi,i}^{\downarrow}}{L+1-j}<p_{\psi,j-1}^{\downarrow}\right\}.\nonumber
\end{eqnarray}
Then, there exists a suitable LOCC map to transform $\psi$ to $\eta_{\psi,L}$, and we can get the following equation:
\begin{eqnarray}\label{Ceq}
\max_{C:LOCC} F(C(\psi),\Phi_L)=F(\eta_{\psi,L},\Phi_L)
\end{eqnarray} 
where $C:\mathcal{S}(\mathcal{H}_A\otimes\mathcal{H}_B)\to\mathcal{S}(\mathcal{H}'_A\otimes\mathcal{H}'_B)$ runs over LOCC maps. Similarly, when we define the pure state $\zeta_{\psi,L}$ in $\mathcal{H}_A\otimes\mathcal{H}_B$ as 
\begin{eqnarray}
\zeta_{\psi,L}= 
\sqrt{\sum_{i=1}^{L}p_{\psi,i}^{\downarrow}}^{~-1}\displaystyle\sum_{i=1}^{L}\sqrt{p_{\psi,i}^{\downarrow}}|i\rangle |i\rangle, 
\nonumber
\end{eqnarray}
 there exists a suitable LOCC map to transform $\Phi_L$ to $\zeta_{\psi,L}$, and the following equation holds as is shown in \cite{VJN}:
\begin{eqnarray}\label{Deq}
\displaystyle\max_{D:LOCC} F(\psi,D(\Phi_L))
=F(\psi,\zeta_{\psi,L})=\sqrt{\sum_{i=1}^{L}p_{\psi,i}^{\downarrow}}
\end{eqnarray}
where $D:\mathcal{S}(\mathcal{H}'_A\otimes\mathcal{H}'_B)\to\mathcal{S}(\mathcal{H}_A\otimes\mathcal{H}_B)$ run over LOCC maps. Moreover, we easily get the equation 
\begin{eqnarray}\label{minequation}
\displaystyle\max_{C,D:LOCC}F(\psi, D\circ C(\psi))
=\displaystyle\max_{D:LOCC} F(\psi,D(\Phi_L)),
\end{eqnarray}
where $C:\mathcal{S}(\mathcal{H}_A\otimes\mathcal{H}_B)\to\mathcal{S}(\mathcal{H}'_A\otimes\mathcal{H}'_B)$ and $D:\mathcal{S}(\mathcal{H}'_A\otimes\mathcal{H}'_B)\to\mathcal{S}(\mathcal{H}_A\otimes\mathcal{H}_B)$ run over LOCC maps.

~

\noindent[Proof of (\ref{eq})]~Due to (\ref{minequation}),
\begin{eqnarray}
\delta_n(\psi,M_n)
\ge \min_{L_n\in\Z_{>0}}  d(\psi^{\otimes n}\to\Phi^{\otimes {L_n}})+d(\Phi^{\otimes {L_n}}\to\psi^{\otimes M_n})\nonumber
\end{eqnarray}
holds without any condition for $M_n$. Next, we prove the converse inequality for $M_n\le n$. Let us fix an arbitrary $L_n\in\Z_{>0}$. Since there exists a suitable LOCC map from $\eta_{\psi^{\otimes n},2^{L_n}}$ to $\zeta_{\psi^{\otimes M_n},2^{L_n}}$ when $M_n$ is less than or equal to $n$, we get 
\begin{eqnarray}\label{inequality}
&&\delta_n(\psi,M_n)\nonumber\\
&&\le d(\eta_{\psi^{\otimes n},2^{L_n}},\Phi^{\otimes L_n})+d(\psi^{\otimes M_n},\zeta_{\psi^{\otimes M_n},2^{L_n}})\nonumber\\
&&\le d(\psi^{\otimes n}\to\Phi^{\otimes {L_n}})+d(\Phi^{\otimes {L_n}}\to\psi^{\otimes M_n}).
\end{eqnarray}
Here, we used (\ref{Ceq}) and (\ref{Deq}) to show the inequality (\ref{inequality}). $\blacksquare$

~

\noindent[Proof of Theorem \ref{2-order}]~
We introduce the following values for a sequence $\overline{\rho}=\{\rho_n\}_n$ of general quantum states.
\begin{eqnarray}
&\overline{K}(a,b|\overline{\rho}):=\overline{\rm lim}{\rm Tr}\rho_n\{-{\rm log}\rho_n\le an+b\sqrt{n}\}&\nonumber\\
&\underline{K}(a,b|\overline{\rho}):=\underline{\rm lim}{\rm Tr}\rho_n\{-{\rm log}\rho_n\le an+b\sqrt{n}\}&\nonumber
\end{eqnarray}
When $\psi_n$ is an arbitrary entangled pure state on a composite system $\mathcal{H}_{A,n}\otimes\mathcal{H}_{B,n}$, the following inequalities hold for a sequence $\mathrm{Tr}_B\overline{\psi}:=\{\mathrm{Tr}_B\psi_n\}_n$.
\begin{eqnarray}
1-\hspace{-0.3em}\displaystyle\lim_{\gamma\to+0}\overline{K}(a,b+\gamma|{\rm Tr}_{B}\overline{\psi})
\le\underline{\rm lim}d(\Phi^{\otimes an+b\sqrt{n}} \to \psi_n)~~\label{r-1}
\end{eqnarray}

\vspace{-2em}
\begin{eqnarray}
\overline{\rm lim}d(\Phi^{\otimes an+b\sqrt{n}} \to \psi_n)
\le1-\underline{K}(a,b|{\rm Tr}_{B}\overline{\psi})\label{r-2}
\end{eqnarray}

\vspace{-2em}
\begin{eqnarray}
\hspace{0em}\displaystyle\lim_{\gamma\to+0}\underline{K}(a,b-\gamma|{\rm Tr}_{B}\overline{\psi})
\le\underline{\rm lim}d(\psi_n \to \Phi^{\otimes an+b\sqrt{n}})\label{s-1}
\end{eqnarray}

\vspace{-2.0em}
\begin{eqnarray}
\overline{\rm lim}d(\psi_n \to \Phi^{\otimes an+b\sqrt{n}})
\le\displaystyle\lim_{\gamma\to+0}\overline{K}(a,b+\gamma|{\rm Tr}_{B}\overline{\psi})\label{s-2}
\end{eqnarray}

At first, we prove (\ref{r-1}). By (\ref{Deq}), for an arbitrary state $\psi_n$, arbitrary positive integers ${L_n},{L'_n}$, and an arbitrary entanglement dilution transformation $({L_n},D_n)$, the inequality $F^2(\psi_n,D_n(\Phi^{\otimes L_n}))\le\mathrm{Tr}\rho\{\rho\ge\frac{1}{2^{L'_n}}\}+\frac{2^{L_n}}{2^{L'_n}}$ holds. When ${L_n}={an+b\sqrt{n}}$, ${L'_n}={an+(b+\gamma)\sqrt{n}}$ in the inequality, we get (\ref{r-1}) by taking $\underline{\rm lim}$ and ${\rm lim}_{\gamma\to+0}$.

Next, we prove (\ref{r-2}). By (\ref{Deq}), for an arbitrary state $\psi_n$ and an arbitrary positive integer ${L_n}$, there is an entanglement dilution transformation $({L_n},D_n)$ satisfying the inequality $\mathrm{Tr}\rho\{\rho\ge\frac{1}{2^{L_n}}\}\le F^2(\psi_n,D_n(\Phi^{\otimes L_n}))$. When ${L_n}={an+b\sqrt{n}}$ in the inequality, we get (\ref{r-2}) by taking $\underline{\rm lim}$.

Next, we prove (\ref{s-1}). By Lemmas 4 and 5 in \cite{Hay2}, for an arbitrary state $\psi_n$, arbitrary positive integers ${L_n}\ge {L'_n}$, and an arbitrary concentration map $({L_n},C_n)$, the inequality 
\begin{eqnarray}
&&F^2(C_n(\psi_n),\Phi^{\otimes L_n})\nonumber\\
&\le&\frac{1}{2^{L_n}}\Big(\sqrt{\mathrm{Tr}\{\rho_n\ge{1}/{2^{L'_n}}\}}\sqrt{\mathrm{Tr}\rho\{\rho_n\ge{1}/{2^{L'_n}}\}}\\
&&\hspace{-0.5em}+\sqrt{M-\mathrm{Tr}\{\rho_n\ge{1}/{2^{L'_n}}\}} \sqrt{1-\mathrm{Tr}\rho\{\rho_n\ge{1}/{2^{L'_n}}\}}\Big)^2 \nonumber
\end{eqnarray}
holds. When ${L_n}={an+b\sqrt{n}}$, ${L'_n}={an+(b-\gamma)\sqrt{n}}$ in the inequality, we get (\ref{s-1}) by taking $\underline{\rm lim}$ and ${\rm lim}_{\gamma\to+0}$.

Finally, we prove (\ref{s-2}). It is enough to prove $\overline{\rm lim}d(\psi_n \to \Phi^{\otimes an+b\sqrt{n}}) \le \overline{K}(a,b+\gamma|{\rm Tr}_{B}\overline{\psi})$ for an arbitrary positive real number $\gamma$. When $\overline{K}(a,b+\gamma|{\rm Tr}_{B}\overline{\psi})=1$, the inequality is obvious. Thus, we assume $\overline{K}(a,b+\gamma|{\rm Tr}_{B}\overline{\psi})<1$. By Lemma 9 and (1) in \cite{Hay2}, for an arbitrary state $\psi_n$ and an arbitrary positive integer ${L_n}$, there is a concentration map $({L_n},C_n)$ satisfying the inequality $1-\mathrm{Tr}\rho_n\{\rho_n\ge x_{2^{L_n}}\}\le F^2(C_n(\psi_n),\Phi^{\otimes L_n})$ where let $h_n(x)$ be $\mathrm{Tr}(\rho_n-x)\{\rho_n-x\ge0\}$ and $x_{{L_n}}$ satisfy $\left\lfloor\frac{1}{x_{{L_n}}}(1-h_n(x_{{L_n}}))\right\rfloor=2^{L_n}$. For ${L_n}:={an+(b+\gamma)\sqrt{n}}+\mathrm{log}(1-h_n(2^{-an-(b+\gamma)\sqrt{n}}))$ and ${L'_n}:={an+b\sqrt{n}}$, we can take as $x_{{L_n}}=2^{-an-(b+\gamma)\sqrt{n}}$. Since $\overline{\lim}h_n(2^{-an-(b+\gamma)\sqrt{n}})\le\overline{K}(a,b+\gamma|\overline{\rho})<1$, ${L'_n}<{L_n}$ holds for enough large integer $n$. Therefore, 
\begin{eqnarray}\label{inequality10}
1-\mathrm{Tr}\rho_n\{\rho_n\ge x_{{L_n}}\}
&\le& F^2(C_n(\psi_n),\Phi^{\otimes L_n}) \nonumber\\
&\le& F^2(E'_n(\psi_n),\Phi^{\otimes{L'_n}})
\end{eqnarray}
holds for suitable $({L_n},C_n)$ and $({L'_n},E'_n)$ for enough large integer $n$. By taking $\underline{\lim}$ in the inequality (\ref{inequality10}), we get $\overline{\rm lim}d(\psi^{\otimes n} \to \Phi^{\otimes an+b\sqrt{n}}) \le \overline{K}(a,b+\gamma|{\rm Tr}_{B}\overline{\psi})$.

Let $\psi$ be a pure state in $\mathcal{H}_A\otimes\mathcal{H}_B$ except maximally entangled state. Then, $V(\mathrm{Tr}_B \psi)$ is not $0$. 
When $\mathcal{H}_{A,n}=\mathcal{H}_{A}^{\otimes n}$, $\mathcal{H}_{B,n}=\mathcal{H}_{B}^{\otimes n}$, and $\psi_n=\psi^{\otimes n+b'\sqrt{n}}$,
\begin{eqnarray}
&&\overline{K}(a,b|{\rm Tr}_{B}\overline{\psi})=\underline{K}(a,b|{\rm Tr}_{B}\overline{\psi})\nonumber\\
&&=\left\{
\begin{array}{ll}
0 & \mathrm{if}~a<H(\mathrm{Tr}_B{\psi}) \\
G\left(\frac{b-H(\mathrm{Tr}_B{\psi})b'}{\sqrt{V(\mathrm{Tr}_B{\psi})}}\right) & \mathrm{if}~a=H(\mathrm{Tr}_B{\psi}) \\
1 & \mathrm{if}~a>H(\mathrm{Tr}_B{\psi}) 
\end{array}
\right.
\end{eqnarray}
due to the (classical) central limit theorem. Thus, Theorem \ref{2-order} holds. $\blacksquare$

~

\noindent[Proof of Theorem \ref{MCREineq}]~
We only have to show that $\displaystyle\underline{\lim}\delta_n(\psi) \ge1$ holds for any pure state $\psi$ except maximally entangled states. Let $\delta(\{L_n\})$ be $\underbar{\rm lim}d(\psi^{\otimes n} \to \Phi^{\otimes L_n})+d(\Phi^{\otimes L_n} \to \psi^{\otimes n})$. Then, $\underline{\rm lim}\delta_n(\psi)=\min_{\{L_n\}}\delta(\{L_n\})$ holds by (\ref{eq}). Moreover, we get $\delta(\{L_n\})\ge1$ as follows. We can take a subsequence $n_k$ of $n$ satisfying that $\delta(\{L_n\})={\rm lim}_k d(\psi^{\otimes n_k} \to \Phi^{\otimes L_{n_k}})+d(\Phi^{\otimes L_{n_k}} \to \psi^{\otimes n_k})$ holds, and $a:={\rm lim}_k {L_{n_k}}/{n_k}$ and $b:={\rm lim}_k \sqrt{n_k}({L_{n_k}}/{n_k}-a)$ exist in $[-\infty,+\infty]$ by repeatedly taking the subsequence, if necessary. 

When $a> H(\mathrm{Tr}_B{\psi})$, $\delta(\{L_n\})\ge1$ because $\underline{\rm lim}_k d(\psi^{\otimes n_k} \to \Phi^{\otimes L_{n_k}})=1$ by Theorem \ref{2-order}. Similarly, when $a< H(\mathrm{Tr}_B{\psi})$,  $\delta(\{L_n\})\ge1$ because $\underline{\rm lim}_k d(\Phi^{\otimes L_{n_k}} \to \psi^{\otimes n_k})=1$ by Theorem \ref{2-order}. 
We treat the case when $a=H(\mathrm{Tr}_B{\psi})$. Let $\epsilon$ be an arbitrary positive real number. By the definition of $b$,  $H(\mathrm{Tr}_B{\psi})n_k+(b-\epsilon)\sqrt{n_k} < L_{n_k} <H(\mathrm{Tr}_B{\psi})n_k+(b+\epsilon)\sqrt{n_k}$ holds for large enough $k$. Then, $d(\Phi^{\otimes L_{n_k}}\to\psi^{\otimes n_k})\ge d(\Phi^{\otimes H(\mathrm{Tr}_B{\psi})n_k+(b+\epsilon)\sqrt{n_k}}\to\psi^{\otimes n_k})$ and $d(\psi^{\otimes n_k}\to\Phi^{\otimes L_{n_k}})\ge d(\psi^{\otimes n_k}\to\Phi^{\otimes H(\mathrm{Tr}_B{\psi})n_k+(b-\epsilon)\sqrt{n_k}})$ hold. Therefor, we get $\underline{\lim}_k d(\Phi^{\otimes L_{n_k}}\to\psi^{\otimes n_k})\ge G((b+\epsilon)/\sqrt{V_{\mathrm{Tr}_B\psi}})$ and $\underline{\lim}_k d(\psi^{\otimes n_k}\to\Phi^{\otimes L_{n_k}})\ge 1-G((b-\epsilon)/\sqrt{V_{\mathrm{Tr}_B\psi}})$ by Theorem \ref{2-order}. Thus, $\delta(\{L_n\})\ge1-G((b-\epsilon)/\sqrt{V_{\mathrm{Tr}_B\psi}}) +G((b+\epsilon)/\sqrt{V_{\mathrm{Tr}_B\psi}})$ is derived. Since $\epsilon$ is arbitrary, we get $\delta(\{L_n\})\ge1$. $\blacksquare$


~

\noindent[Proof of Theorem \ref{2nd.rate}]~
We represent $H(\mathrm{Tr}_B\psi)$ and $V(\mathrm{Tr}_B\psi)$ as $H$ and $V$ in this proof. We can restrict $M_n$ to the form ${\alpha n+\beta\sqrt{n}}$ in the same way as Proof of Theorem \ref{MCREineq}. By the condition $\overline{\lim}\delta_n(\psi,{\alpha n+\beta\sqrt{n}})\le\epsilon<1$ and Theorem \ref{2-order}, the first order rate $\alpha$ of $M_n$ is restricted to $H$. Then,
\begin{eqnarray}
&&\lim \delta_n(\psi,{ n+\beta\sqrt{n}})\nonumber\\
&&=\min_{\{L_n\}}\lim d(\psi^{\otimes n}\to\Phi^{\otimes L_n})+d(\Phi^{\otimes L_n}\to\psi^{\otimes n+\beta\sqrt{n}})\nonumber\\
&&=\min_{b\in\R}\lim d(\psi^{\otimes n}\to\Phi^{\otimes Hn+b\sqrt{n}})\nonumber\\
&&~~~~~~~~~~+d(\Phi^{\otimes Hn+b\sqrt{n}}\to\psi^{\otimes n+\beta\sqrt{n}})\nonumber\\
&&=\min_{b\in\R} G\left(\frac{b}{\sqrt{V}}\right)+1-G\left(\frac{b-H\beta}{\sqrt{V}}\right)\nonumber\\
&&=G\left(\frac{H\beta}{2\sqrt{V}}\right)+1- G\left(\frac{-H\beta}{2\sqrt{V}}\right).\nonumber
\end{eqnarray}
Therefore, we get the following equation.
\begin{eqnarray}
&&R(\psi,\epsilon)\nonumber\\
&&=\inf_{\beta}\left\{{\lim}\frac{n-(n+\beta\sqrt{n})}{\sqrt{n}}\Big| {\lim}\delta_n(\psi,n+\beta\sqrt{n})\le\epsilon\right\}\nonumber\\
&&=\inf_{\beta}\Big\{-\beta \Big| G\left(\frac{H\beta}{2\sqrt{V}}\right)+1- G\left(\frac{-H\beta}{2\sqrt{V}}\right)\le\epsilon \Big\}\nonumber\\
&&=\frac{2\sqrt{V}}{H}G^{-1}\left(1-\frac{\epsilon}{2}\right),\nonumber
\end{eqnarray}
where $G^{-1}$ means the inverse function of the Gaussian distribution function $G$.~$\blacksquare$


\end{document}